\documentclass[11pt,a4paper,twocolumn]{article}
\usepackage{jheppub}
\usepackage{amsmath}
\usepackage{amssymb}
\usepackage{graphicx,caption}
\usepackage{subcaption}
\usepackage{float}
\usepackage{verbatim}
\usepackage{hyperref}							 
\usepackage[font=footnotesize]{caption}
\usepackage{slashed}
\usepackage{siunitx}
\usepackage{nicefrac}
\usepackage{tabularx}
\usepackage{tabulary}
\usepackage{pstricks}
\usepackage{color}
\usepackage{subfiles}
\usepackage{ stmaryrd }
\usepackage[labelfont=bf]{caption}
\usepackage{cleveref}
\usepackage{bbm}
\usepackage{amsmath,amsfonts,braket,slashed,amssymb,tikz,bm,psfrag,graphicx,color,dsfont}
\usepackage{xspace}
 
\usepackage{hepparticles}

\newcommand{\mttwo}{m_{T2}}

\newcommand{\hc}[1]{{}^\ddagger #1}
\newcommand*{\tmp}[4]{\ensuremath{
    {#4
    \ifx\empty#3\empty\ifx\empty#1\empty\else^{#1}\fi\else^{#1(#3)}\fi
    \ifx\empty#2\empty\else_{#2}\fi}
}}
\newcommand*{\qq }[4][]{\tmp{#2}{#3}{#4}{#1{O}}}

\newcommand*{\eftOp}[4]{\ensuremath{
    {#4
    \ifx\empty#3\empty\ifx\empty#1\empty\else^{#1}\fi\else^{#1(#3)}\fi
    \ifx\empty#2\empty\else_{#2}\fi}
}}

\newcommand{\FDFI}{(\varphi^\dagger i\!\!\overleftrightarrow{D}^I_\mu\varphi)}

\newcommand{\cpQa} {\eftOp{3}{\varphi Q}{}{c}\xspace}
\newcommand{\cptb} {\eftOp{}{\varphi t  b}{}{c}\xspace}

\newcommand{\ctW}  {\eftOp{}{ tW}{}{c}\xspace}
\newcommand{\cbW}  {\eftOp{}{ bW}{}{c}\xspace}
\newcommand{\ctG}  {\eftOp{}{ t G}{}{c}\xspace}

\newcommand{\cQla} {\eftOp{3}{Q\ell}{\ell}{c}\xspace}
\newcommand{\cQlaa} {\eftOp{3}{Q\ell}{e}{c}\xspace}
\newcommand{\cQlbb} {\eftOp{3}{Q\ell}{\mu}{c}\xspace}
\newcommand{\cQlcc} {\eftOp{3}{Q\ell}{\tau}{c}\xspace}
\newcommand{\ctlS} {\eftOp{S}{ t}{\ell}{c}\xspace}
\newcommand{\ctlT} {\eftOp{T}{t}{\ell}{c}\xspace}

\newcommand{\emu}{\ensuremath{e^{\pm}\mu^{\mp}}\xspace}
\newcommand{\mumu}{\ensuremath{\mu^{\pm}\mu^{\mp}}\xspace}
\newcommand{\ee}{\ensuremath{e^{\pm}e^{\mp}}\xspace}

\begin{document}
\begin{titlepage}
	
	\begin{flushright}
		\normalsize
		\today
	\end{flushright}
	
	\vspace{1.0cm}
	\begin{center}
		\Large\bf\boldmath 
		Probing effective operators in single top quark production in association with a lepton-neutrino pair
	\end{center}
	
	\vspace{0.5cm}
	\begin{center}
		Reza Goldouzian\footnote{Rgoldouz@nd.edu} and Michael D. Hildreth\footnote{Mikeh@nd.edu}  \\
		\vspace{0.7cm} 
		{\sl Department of Physics, 225 Nieuwland Science Hall,University of Notre Dame, Notre Dame, IN 46556, USA}
	\end{center}
	
	\vspace{0.8cm}
	\begin{abstract}
		
  We study single top quark production in association with a lepton-neutrino pair at the LHC  within the framework of the Standard Model Effective Field Theory (SMEFT). We focus on relevant two-quark-two-lepton operators (\qq{3}{lq}{}, \qq{1}{lequ}{}, and \qq{3}{lequ}{}). It is known that these operators  have tiny  effects on the inclusive cross section of standard model tW production and are thus typically ignored in the SMEFT searches. However, we show that by employing smart observables, such as $m_{T2}$, the pp$\rightarrow$t$\ell\nu$ process is significantly sensitive to these operators. We set the most stringent limit on the coupling strength of the \qq{3}{lq}{} operator by reinterpreting the results of a search for new phenomena  with two opposite-charge leptons, jets and missing transverse momentum at $\sqrt{s}$ = 13 TeV performed by the ATLAS collaboration. The limits derived on the \qq{1}{lequ}{} and \qq{3}{lequ}{} operators are  comparable to those obtained  from probing EFT effects in pp$\rightarrow$t$\Bar{\rm t}\ell\nu$ and pp$\rightarrow$t$\Bar{\rm t}\ell\ell$ processes at the LHC. Consequently, we propose to include the effects of these three operators on the pp$\rightarrow$t$\ell\nu$ process in future global SMEFT analyses to increase the sensitivity and to reduce possible degeneracies.  
	\end{abstract}
	
\end{titlepage}


	\section{Introduction}
All direct searches at the large hadron collider (LHC)  have so far been unsuccessful in finding clear indications of physics beyond the standard model (BSM). This might mean that the BSM energy scale ($\Lambda$) is above what can be directly probed at the LHC. If $\Lambda$ is well separated from the electroweak scale, heavy states can be integrated out and then cast into extensions of the SM Lagrangian by higher dimensional operators ($\cal O$) \cite{Weinberg:1978kz}, i.e. operators of dimension D$>$4:
\begin{equation*}
\label{eq:eft}
{\cal{L}}={\cal{L}}_{\text{SM}} + \sum_{D,i} \frac{C_i^D}{\Lambda^{D-4}}{\cal{O}}_i^D\,,
\end{equation*}
The dimensionless effective couplings of new physics, called Wilson coefficients (WC), are denoted by $C_i^D$. 
The structure of the operators is entirely dictated  by the field content and symmetry restrictions. Writing $\cal{L}$ including the SM gauge symmetries and particle content results in the  SM Effective Field Theory (SMEFT) representation~\cite{Brivio:2017vri}.
The SMEFT is a consistent model-independent framework for systematically parameterising and predicting deviations from the SM. Furthermore, 
the SMEFT is a renormalizable theory and its predictions can be improved by adding terms of higher order quantum electrodynamic (QCD) and electroweak corrections~\cite{Degrande:2020evl}. 

For dimension D=6, there are 59 independent operators with minimal flavour violation and Baryon number conservation \cite{Grzadkowski:2010es}. 
Each operator may affect multiple SM physics processes in different parts of the available phase space.
In order to explore the SMEFT parameter space, it is necessary to discover regions sensitive to SMEFT effects for different types of processes
and include them within a coherent global SMEFT analysis.   
In recent years, the SMEFT has been probed in different sectors by measurements of quantities related to electroweak bosons, the Higgs boson and the top quark by performing global fits \cite{Ellis:2018gqa,Buckley:2015nca,Buckley:2015lku,Castro:2016jjv,Hartland:2019bjb,Brivio:2019ius}. 
All these global fits are performed on a few hand-picked unfolded kinematic  distributions provided by the experiments.

Recently, the CMS Collaboration has performed  the most global detector-level SMEFT search to date in the production of one or more top quarks with additional leptons, jets, and b jets  in proton-proton (pp) collisions~\cite{CMS:2023xyc}.
Constraints are set on 26 WCs which affect the pp$\rightarrow$t$\Bar{\rm t}$X and  pp$\rightarrow$tqX (X= W, Z and Higgs bosons and q=light jet) processes.
They have also included non-Z/W/H mediated t$\Bar{\rm t}\ell\ell$ and t$\Bar{\rm t}\ell\nu$ processes for probing the two-quark-two-lepton operators in multilepton final states \cite{Aguilar-Saavedra:2018ksv}.  In this analysis there are two important improvements with respect to the previous version \cite{CMS:2020lrr}:  first, including 10 more sensitive operators in the global fit, and second, employing smarter observables in various regions of the phase space. Consequently,  a more realistic global fit is performed and the $2\sigma$ profiled confidence intervals are improved by factors of approximately 2 to 6, depending on the WC.  
Therefore,  recognizing  sensitive operators in any available phase space and including them in the global fit is a crucial task~\cite{Brehmer:2018eca,Faham:2021zet,Aoude:2022deh,Maltoni:2019aot}.

In this work, we focus on the SMEFT effects on single top quark production in association with a W boson at the LHC. 
We show that the number of sensitive operators increases if we extend the pp$\rightarrow$tW process to include pp$\rightarrow$t$l\nu$.  
We propose a unique phase space for probing the extended  operators and examine its sensitivity by reinterpreting the results of a search for  new phenomena  with two opposite-charge leptons, jets and missing transverse momentum performed by the ATLAS Collaboration \cite{ATLAS:2021hza}. 

\section{SMEFT operators}

In this section, we review the SMEFT operators that affect  single top quark production in association with a lepton-neutrino pair in pp collisions. 
Eight operators  can affect the pp$\rightarrow$t$l\nu$ process at leading order (LO); these operators are listed in Table~\ref{tab:eftOperators} together with corresponding WCs \cite{Aguilar-Saavedra:2018ksv, Zhang:2010dr}. The operators are classified into two categories: operators that involve two quarks and gauge bosons, and operators that involve two quarks and two leptons. 

\begingroup
\renewcommand{\arraystretch}{1.4} 
\begin{table}[hbt!]
\caption{List of operators that have effects on the pp$\rightarrow$t$l\nu$ process. The $ijkl$ are flavor indices; $q$  and $\ell$ are the left-handed fermion doublets;  $u$, $d$, and $e$ are the right-handed fermion singlets; $\varphi$ is the Higgs doublet;  $\tau^I$ are the Pauli matrices; $\varepsilon\equiv i\tau^2$; $T^A\equiv \lambda^A/2$ where $\lambda^A$ are Gell-Mann matrices \cite{Aguilar-Saavedra:2018ksv}.}
\begin{center}
\begin{tabular}{lll}
\hline
Operator                     & Definition                                                             & WC                       \\ \hline
\multicolumn{3}{c}{Two-quark operators} \\ \hline
$\qq{3}{\varphi q}{ij}$           &$ \FDFI (\bar{q}_i\gamma^\mu\tau^I q_j) $                     & $ \cpQa  $    \\ 
$\hc{\qq{}{\varphi ud}{ij}}$   & $(\tilde\varphi^\dagger iD_\mu\varphi) (\bar{u}_i\gamma^\mu d_j)$      & $ \cptb$   \\ 
$\hc{\qq{}{dW}{ij}}$            & $(\bar{q}_i\sigma^{\mu\nu}\tau^Id_j)\:{\varphi} W_{\mu\nu}^I$       & $ \cbW$    \\ 
$\hc{\qq{}{uW}{ij}}$            & $(\bar{q}_i\sigma^{\mu\nu}\tau^Iu_j)\:\tilde{\varphi}W_{\mu\nu}^I$  & $ \ctW$   \\ 
$\hc{\qq{}{uG}{ij}}$            & $(\bar{q}_i\sigma^{\mu\nu}T^Au_j)\:\tilde{\varphi}G_{\mu\nu}^A$       & $\ctG$    \\ \hline
\multicolumn{3}{c}{Two-quark-two-lepton operators} \\ \hline
$\qq{3}{lq}{ijkl}$                &$(\bar l_i\gamma^\mu \tau^I l_j) (\bar q_k\gamma^\mu \tau^I q_l)$          & $ \cQla $ \\ 
$\hc{\qq{1}{lequ}{ijkl}}$   &$(\bar l_i e_j)\;\varepsilon\;(\bar q_k u_l)$                                  & $ \ctlS$        \\ 
$\hc{\qq{3}{lequ}{ijkl}}$    &$(\bar l_i \sigma^{\mu\nu} e_j)\;\varepsilon\;(\bar q_k \sigma_{\mu\nu} u_l)$  & $ \ctlT$ \\ \hline
\end{tabular}
\label{tab:eftOperators}
\end{center}
\end{table}
\endgroup

In the SM, the pp$\rightarrow$t$l\nu$  process involves tW production with the leptonic decays of the W boson.  The   $\qq{3}{\varphi q}{ij}$, $\hc{\qq{}{\varphi ud}{ij}}$,  $\hc{\qq{}{dW}{ij}}$, and $\hc{\qq{}{uW}{ij}}$ operators modify the SM Wtb vertex and thus affect the tW production\cite{Aguilar-Saavedra:2008quj}.  The $\qq{3}{\varphi q}{ij}$ operator has the same  structure as the SM Wtb vertex and only rescales the Wtb coupling without affecting any kinematic distribution.
In the $m_b=0$ limit, the $\hc{\qq{}{\varphi ud}{ij}}$ and $\hc{\qq{}{dW}{ij}}$ operators do not interfere with the SM amplitude and only contribute at order $1/\Lambda^4$ to the cross section.  The  $\hc{\qq{}{uW}{ij}}$ operator involves the right-handed top quark and interferes with the SM amplitudes. The $\hc{\qq{}{\varphi ud}{ij}}$,  $\hc{\qq{}{dW}{ij}}$, and $\hc{\qq{}{uW}{ij}}$ operators affect the  W boson polarization fractions in top quark decays and are well constrained via the W polarization measurements by the CMS and ATLAS Collaborations \cite{CMS:2020ezf,Jueid:2018wnj}.
These operators can also affect the pp$\rightarrow$t$\Bar{\rm t}$X and pp$\rightarrow$tqX processes \cite{CMS:2023xyc, Zhang:2016omx}.
The $\hc{\qq{}{uG}{ij}}$ operator  contributes to tW production via modification of the SM top-gluon vertex \cite{Zhang:2010dr}.  This operator is also well constrained using t$\Bar{\rm t}$ events at the LHC \cite{CMS:2019nrx,CMS:2019zct}.

The  $\qq{3}{lq}{ijkl}$,  $\hc{\qq{1}{lequ}{ijkl}}$, and $\hc{\qq{3}{lequ}{ijkl}}$ operators do not have a W boson  and directly connect a top-bottom quark pair to a lepton-neutrino pair \cite{Stolarski:2020cvf}. 
The  $\qq{3}{lq}{ijkl}$,  $\hc{\qq{1}{lequ}{ijkl}}$, and $\hc{\qq{3}{lequ}{ijkl}}$  operators have vector-, scalar-, and tensor-like Lorentz structure, respectively.
It is worth mentioning that these operators are of great interest for new physics searches because of the observed anomalies in B meson decays \cite{Cornella:2021sby, CMS:2022ztx}.  They can   represent  the low energy  effects of heavy new physics particles such as W', Z', leptoquarks, etc. which could be solutions that can describe the ``B anomalies"\cite{Blanke:2022deg}. These two-quark-two-lepton operators with a top quark leg also affect the t$\Bar{\rm t}\ell\ell$ and t$\Bar{\rm t}\ell\nu$ processes and are probed in three lepton final states by the CMS collaboration \cite{CMS:2023xyc}.

\begin{figure*}[tb]
\begin{center}
      \includegraphics[width=2.0\columnwidth]{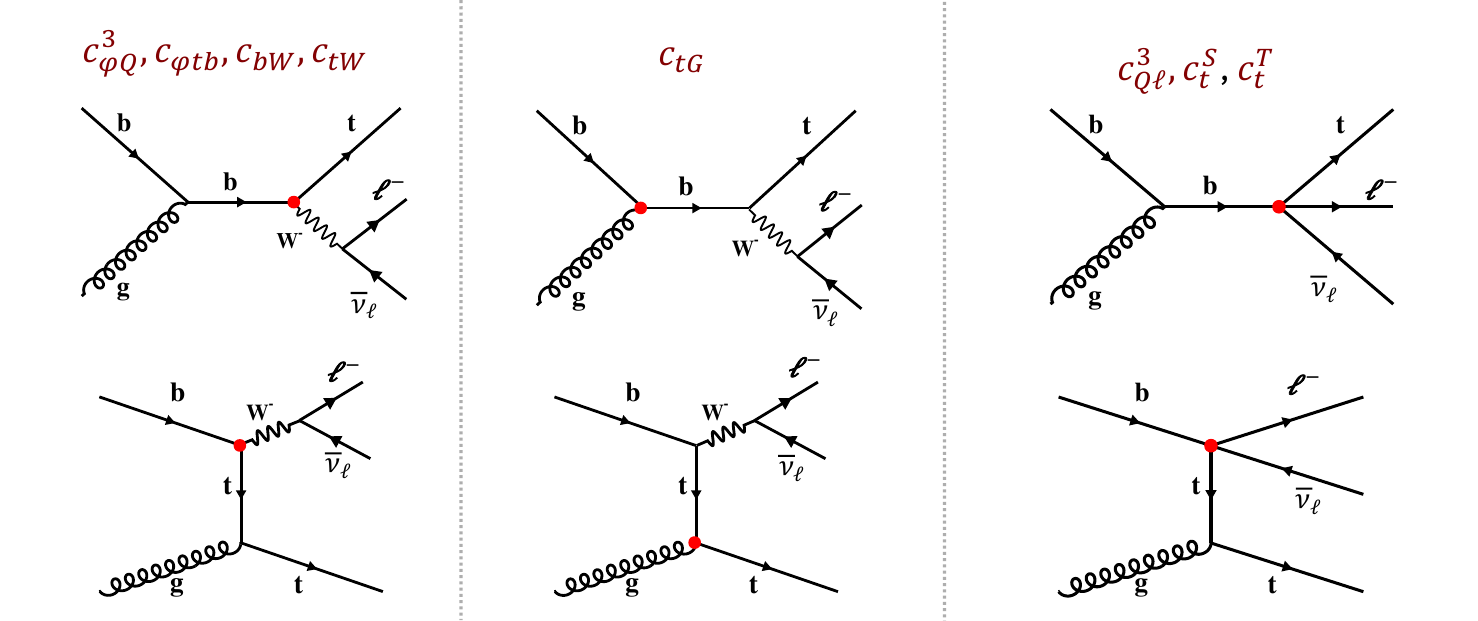}
    \caption{Representative Feynman diagrams for single top quark production in association with a lepton-neutrino pair via SMEFT interactions. The red circles mark the EFT vertices.
    \label{Feyn}}
    \end{center}
\end{figure*}

\section{Signal sample simulation}
We aim to study dimension-six EFT effects on  t$l\nu$ production at the LHC.
Representative Feynman diagrams for  pp$\rightarrow$t$l\nu$ process in the presence of the EFT vertices are shown in Figure. \ref{Feyn}.

The signal sample is modeled at leading order using the Madgraph@NLO Monte Carlo event generator in the five-flavour scheme  \cite{Alwall:2014hca}.
The SMEFT Lagrangian is included using the \textsc{dim6top} model described in Ref.~\cite{Aguilar-Saavedra:2018ksv}.
Events are generated using the  NNPDF3.0 parton distribution function (PDF) set \cite{Ball:2014uwa} and  are passed to PYTHIA8 \cite{Sjostrand:2007gs} for showering and hadronization. 

In the \textsc{dim6top} model, only operators involving one or more top quarks are provided with no limitations on the lepton flavors.  
In this study we assume the same couplings for all generations of leptons  (e.g.  \cQlaa = \cQlbb = \cQlcc). 
All eight operators mentioned in Table~ \ref{tab:eftOperators}  are included simultaneously in the signal sample generation.
Only diagrams with one EFT vertex are included together with the SM diagrams. 
The signal sample is generated at a non-zero point of eight-dimensional WC phase space.
In order to predict the EFT effects at arbitrary points of the WC phase space,  we use the Madgraph@NLO event reweighting technique \cite{Mattelaer:2016gcx}.
By knowing the fact that the cross section of the pp$\rightarrow$t$l\nu$ process depends  quadratically on the WCs and by having  enough event weights,  we can determine the eight-dimensional quadratic function of the cross section in any region of the phase space or observable bin \cite{Schoefbeck:2895759}.

\section{Sensitive  observables}

\begin{figure*}[tb]
\begin{center}
      \includegraphics[width=1.2\columnwidth]{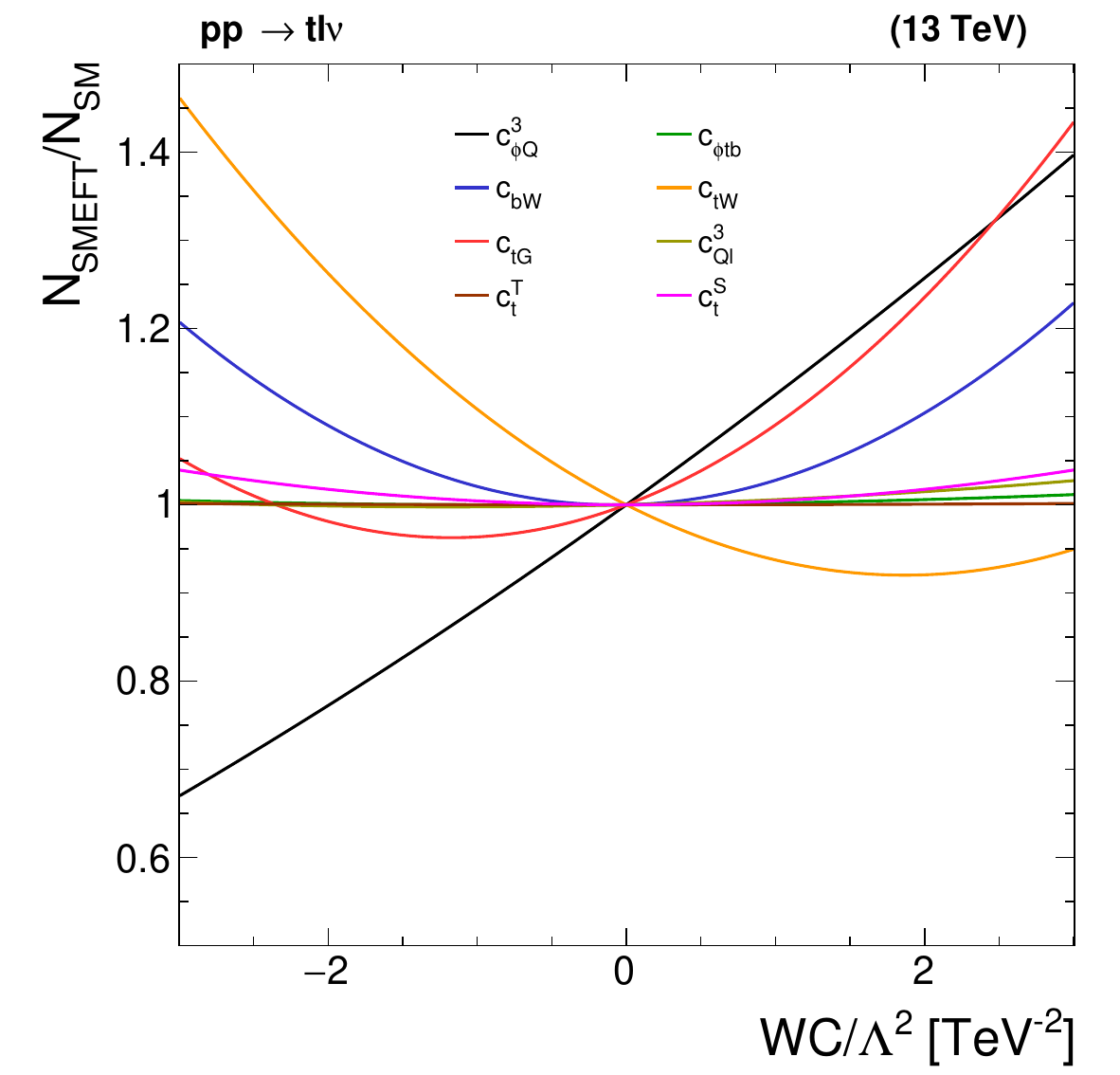}
    \caption{Relative SMEFT  contributions to the  pp$\rightarrow$t$l\nu$ inclusive cross section as a function of the  individual WCs.
    \label{fig:incXS}}
    \end{center}
\end{figure*}

In Figure \ref{fig:incXS},  relative SMEFT  contributions to the  pp$\rightarrow$t$l\nu$ inclusive cross section are shown for individual WCs.
It can be seen that the $\qq{3}{\varphi q}{ij}$,  $\hc{\qq{}{uW}{ij}}$ and $\hc{\qq{}{uG}{ij}}$ operators, with considerable interference with the SM,  have the largest effects.
The $\hc{\qq{}{dW}{ij}}$ operator, with no interference with the SM,  has the second largest effect.
Finally, the $\hc{\qq{}{\varphi ud}{ij}}$ and  two-lepton-two-quark operators have very small effects on the inclusive pp$\rightarrow$t$l\nu$ cross section.

\begin{figure*}[tb]
\begin{center}
      \includegraphics[width=1\columnwidth]{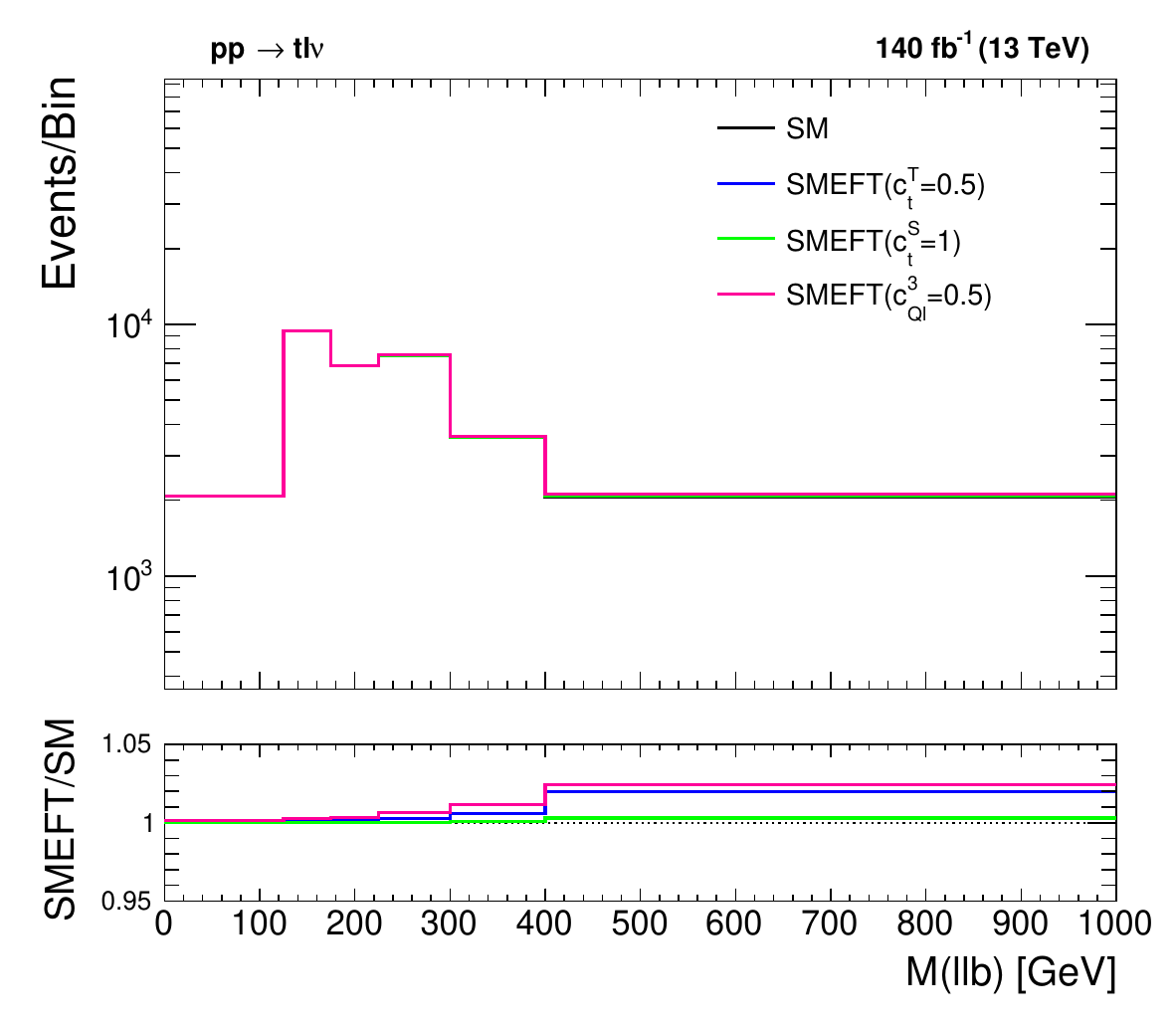}
            \includegraphics[width=1\columnwidth]{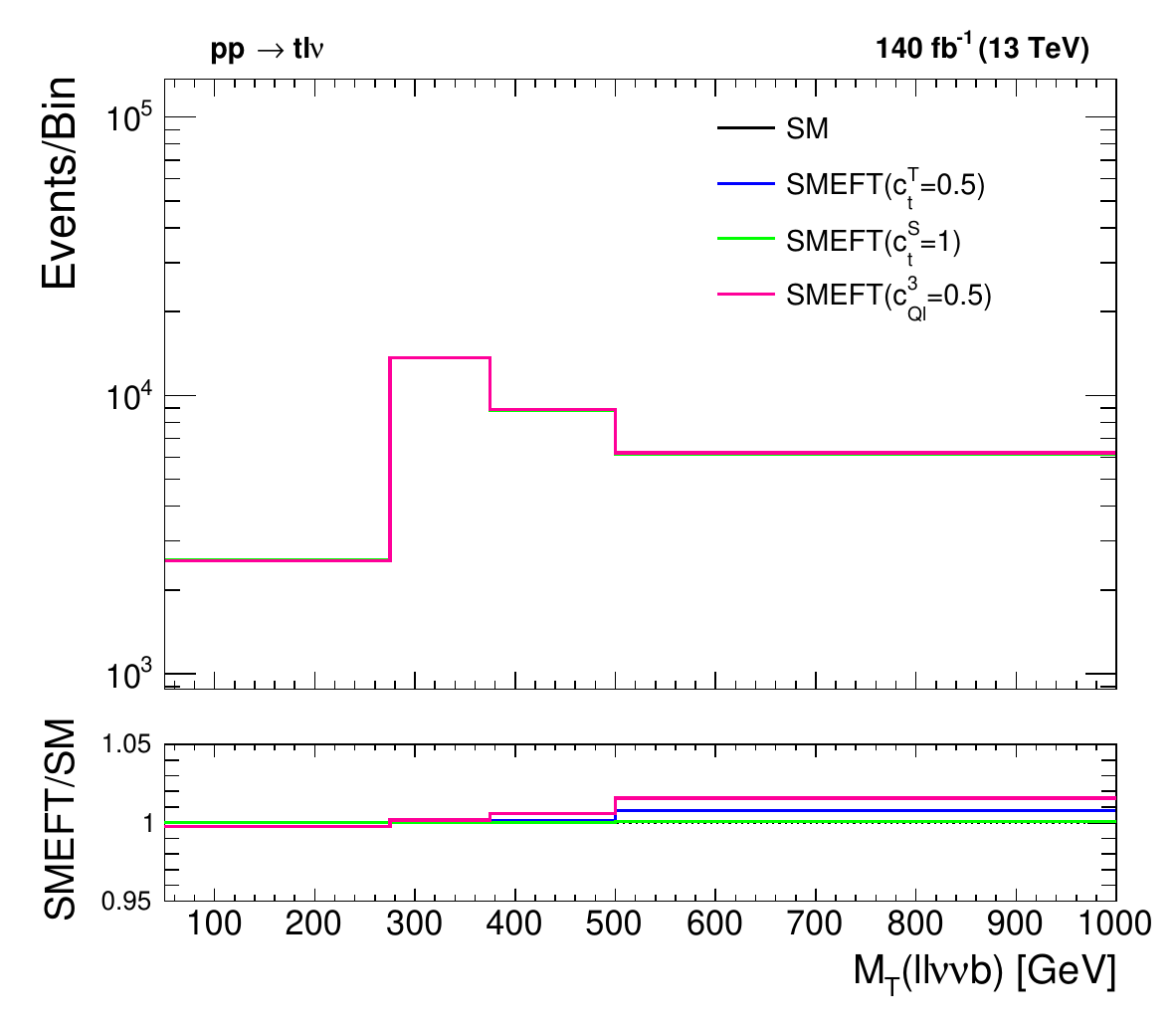}
      \includegraphics[width=1\columnwidth]{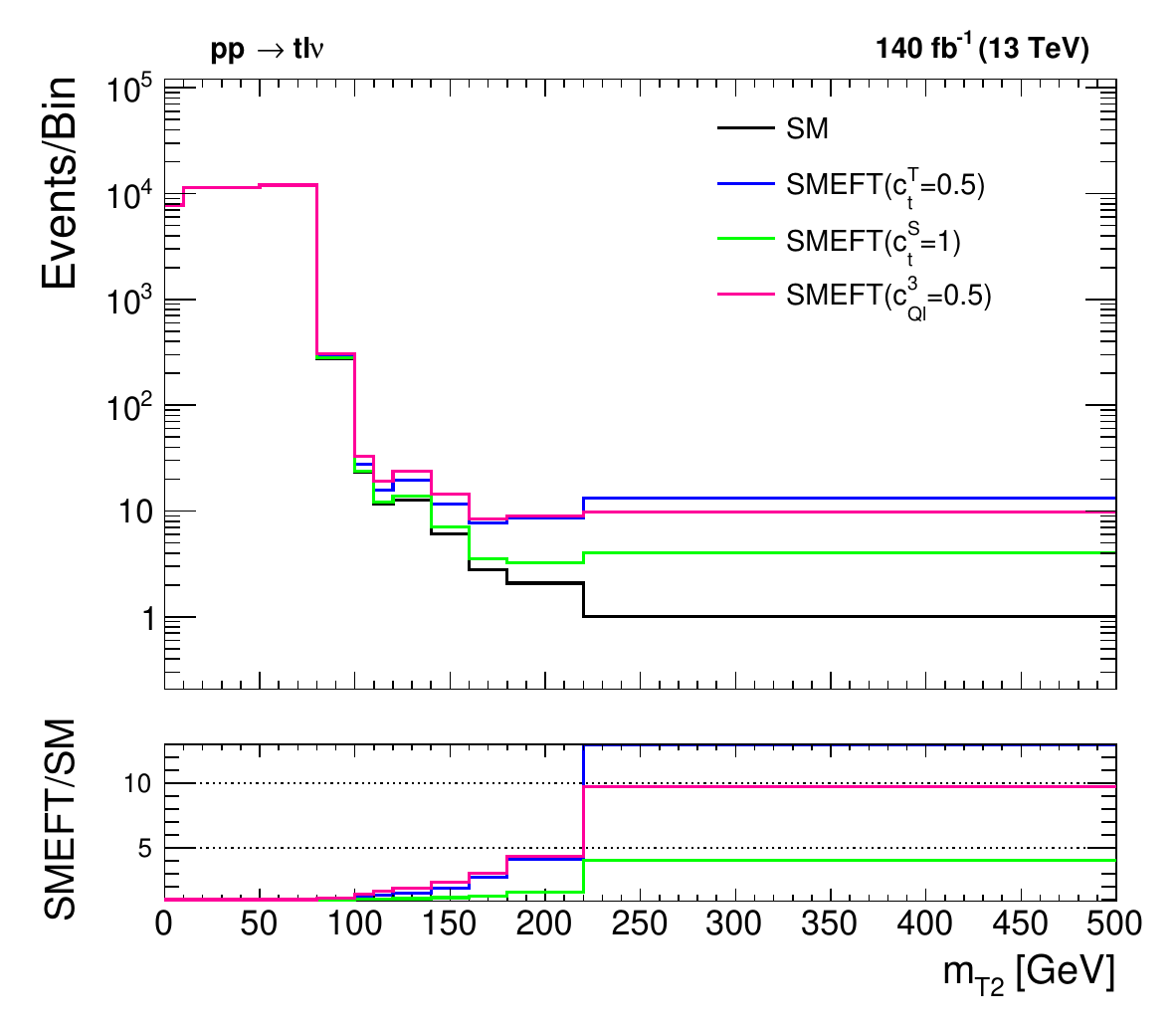}
    \caption{Distributions of the M($\ell\ell b$), M$_T$($ \ell\ell \nu \nu b$), and $m_{T2}$ observables for the pp$\rightarrow$t$l\nu$ process in the SM and SMEFT  with the $\qq{3}{lq}{ijkl}$,  $\hc{\qq{1}{lequ}{ijkl}}$, and $\hc{\qq{3}{lequ}{ijkl}}$ operators. Arbitrary values for the WCs are chosen to show relative contributions.  The lower panel gives the ratio of the SMEFT to SM predictions. Bin sizes for the M($\ell\ell b$) and M($_T \ell\ell \nu \nu b$) distributions are taken from the unfolded distributions in Ref \cite{ATLAS:2017quy}.
    \label{fig:Gen}}
    \end{center}
\end{figure*}

In Ref. \cite{Stolarski:2020cvf}, the authors constrained the two-quark-two-lepton WCs  by reinterpreting the   unfolded differential cross sections  of the pp$\rightarrow$tW$\rightarrow$t$l\nu$ process measured by the ATLAS collaboration \cite{ATLAS:2017quy}.
Although the invariant mass of the $\ell\ell b$ system (M($\ell\ell b$)) and  the transverse mass of the $\ell\ell \nu \nu b$ system (M$_T$($ \ell\ell \nu \nu b$)) show better  separation power between the  EFT and SM events compared to the inclusive cross section,  the 95\% CL limits obtained from these two distributions are much looser than those set by the CMS collaboration by analyzing the  t$\Bar{\rm t}\ell\ell$ and t$\Bar{\rm t}\ell\nu$ processes (see table \ref{tab:resultsCom} for an exact comparison) \cite{CMS:2023xyc}.

There is an important difference between the tl$\nu$ production via the  SM and two-quark-two-lepton operators: the SM process involves on-shell W boson production while the EFT process does not. Therefore, one can discriminate between the SM and EFT events by reconstructing  the mass of the l$\nu$ system. 
In the dilepton final state, the measured missing transverse momentum vector ($ {\overrightarrow{p}}_{\mathrm{T}}^{\mathrm{miss}}$) is associated to the sum of the two neutrinos which causes ambiguity in the reconstruction of the $l\nu$ system.  
A well known variable which can be used to extract information about the mother particle's mass in  events where two massive particles each decay to a detected and an undetected object is $m_{T2}$ \cite{Lester:1999tx}.  
The $m_{T2}$ variable is defined as:

 \begin{align*} \label{mt2eq}
 &  \mttwo (  p_{\mathrm{T,1}},  p_{\mathrm{T,2}},  p_{\mathrm{T}}^{\mathrm{miss}}) = \min_{ q_{\mathrm{T,1}} +  q_{\mathrm{T,2}} =   p_{\mathrm{T}}^{\mathrm{miss}} }  \\ 
& \left\{
  \max [\;
m_{\mathrm{T}}(  p_{\mathrm{T,1}},  q_{\mathrm{T,1}} ),
m_{\mathrm{T}}(  p_{\mathrm{T,2}},  q_{\mathrm{T,2}}
)
\;]
\right\} ,  
\end{align*}

\noindent where $ p_{\mathrm{T}}$ and $  q_{\mathrm{T}}$ are the transverse momenta of the detected  and undetected particles, and
$m_{\mathrm T}$ is the transverse mass 
\begin{equation*}
    m_\textup{T}( p_{\mathrm{T}}, q_{\mathrm{T}} )=\sqrt{2| p_{\mathrm{T}}||  q_{\mathrm{T}} |(1-\mathrm{cos}(\Delta\phi(  p ,  q)))}.
\end{equation*}
The minimisation is performed over all the possible decompositions of $ p_{\mathrm{T}}^{\mathrm{miss}}  =  q_{\mathrm{T,1}} +  q_{\mathrm{T,2}}$.
The $\mttwo$ distribution has an endpoint at the W boson mass for the SM tW events,  while signal events are expected to populate the tails of  $\mttwo$ distribution.
The $m_{T2}$ variable is  used widely in searches for supersymmetric particles \cite{CMS:2017okm,ATLAS:2017mjy} and top quark mass measurements \cite{CDF:2009zjw,CMS:2013wbt}.

In Figure \ref{fig:Gen},  distributions of the M($\ell\ell b$), M$_T$($ \ell\ell \nu \nu b$), and $m_{T2}$ observables are shown for the pp$\rightarrow$t$l\nu$ process in the SM and SMEFT scenarios. Events  with an opposite-sign lepton pair (\ee, \emu, and \mumu) and a b-quark  with $p_{\mathrm{T}}>20$ GeV and $|\eta|<2.4$ are selected at the generator level. We also require events to have  $|  p_{\mathrm{T}}^{\mathrm{miss}}|>40$ GeV to be closer to the selection used in Ref. \citep{Stolarski:2020cvf}. We only show the effects related to non-zero values of the  two-quark-two-lepton WCs.
In the tail of the $m_{T2}$ distribution, EFT variations are enhanced significantly compared to the M($\ell\ell b$) and M$_T$($ \ell\ell \nu \nu b$) distributions. This proves that the   pp$\rightarrow$t$l\nu$ process could be sensitive to the $\qq{3}{lq}{ijkl}$,  $\hc{\qq{1}{lequ}{ijkl}}$, and $\hc{\qq{3}{lequ}{ijkl}}$ operators if one chooses a smart observable.

\section{Experimental inputs}
In order to evaluate the power of the $m_{T2}$ observable for constraining the two-quark-two-lepton operators via tl$\nu$ production, we reinterpret the results of a search  for top squarks in events with two leptons, b-jets and missing transverse energy at $\sqrt{s}=13$ TeV performed by the ATLAS Collaboration  with the full Run-2 data corresponding to 139 fb$^{-1}$~\cite{ATLAS:2021hza}.
Dedicated event selections are optimised in the ATLAS search to probe various decay modes of the top squark. 
One of these event selections, called the ``two-body  selection'', targets the two-body top squark decay into an on-shell top quark and the lightest neutralino. The estimated SM yields in the signal region defined in the ``two-body selection'' are reported in multiple bins of the $m_{T2}$ distribution in the range of [100,inf) GeV.
This signal region is proper for probing the  pp$\rightarrow$t$l\nu$ process  and the $m_{T2}$  variable that is our variable of interest in this study.  Therefore, we use the published results in Ref. ~\cite{ATLAS:2021hza} in the ``two-body  selection'' for constraining the two-quark-two-lepton operators.

In the  ``two-body  selection'',  events are accepted if they have two opposite-sign leptons (\ee, \emu, and \mumu) with the leading lepton  $p_{\mathrm{T}}(\ell 1)$ larger than 25 GeV and the subleading lepton $p_{\mathrm{T}}(\ell 2)$ larger than 20 GeV.  
Electrons (muons) are required to be in the $|\eta|<2.47 (2.40)$ region. Electrons in the transition region between the 
barrel and endcap electromagnetic calorimeters of the ATLAS detector, $1.37<|\eta|<1.52$, are excluded. 
To remove leptons from Drell–Yan and low-mass resonances,  dilepton events with an invariant mass $m_{ll}<20$  are rejected. 
 Events with same flavor (SF) lepton pairs with $71.2<m_{ll}<111.2$ GeV are also rejected to reduce the Z boson background.
Jets are reconstructed using the anti-$k_t$ algorithm \cite{Cacciari:2008gp} with a radius parameter of 0.4 and are selected if they have $p_{\mathrm{T}}>20$ GeV and $|\eta|<2.8$.  Selected jets are tagged as b-jets  using the  MV2C10 boosted decision tree algorithm with 77\% efficiency \cite{ATLAS:2019bwq}. Events should have at least one reconstructed b-tagged jet.
Linked to the missing transverse momentum, the object-based missing transverse momentum’s  significance ($E_{\mathrm{T}}^{\mathrm{miss}} \rm{significance} $) is used  to discriminate
between the events in which the reconstructed $E_{\mathrm{T}}^{\mathrm{miss}}$ originates from weakly interacting particles and the
events in which the $E_{\mathrm{T}}^{\mathrm{miss}}$ is related to the resolution and inefficiencies in particle measurements. 
The $E_{\mathrm{T}}^{\mathrm{miss}} \rm{significance} $ is computed by considering the expected resolution and mismeasurement likelihood of
all the objects that enter the $E_{\mathrm{T}}^{\mathrm{miss}}$  reconstruction, as detailed in Ref.~\cite{ATLAS:2018uid}.
Another variable employed in this search is the azimuthal angle between the  ${\overrightarrow{p}}_{\mathrm{T}}^{\mathrm{miss}}$ and  the vectorial sum of  ${\overrightarrow{p}}_{\mathrm{T}}^{\mathrm{miss}}$ and the leptons' transverse momentum vectors,  denoted as $\Delta\phi_{\rm boost}$. 
Selected events are required to have  $E_{\mathrm{T}}^{\mathrm{miss}}$ significance larger than 12, $\Delta\phi_{\rm boost}<1.5$, and finally $m_{T2}$ greater than 110 GeV.
The two-body selection requirements are summarized in Table \ref{tab:SR}. 

\begin{table}[h]
\caption{\label{tab:SR} Two-body selection}
\begin{center}
\begin{tabular}{l|cc}

Variables                 &  \multicolumn{2}{c}{Cuts} \\ \hline
$p_{\mathrm{T}}(\ell 1)$ [GeV]                 & \multicolumn{2}{c}{$>25$}\\
$p_{\mathrm{T}}(\ell 2)$ [GeV]                 & \multicolumn{2}{c}{$>20$}\\
$m_{ll}$ [GeV]                    & \multicolumn{2}{c}{$>20$}\\
$|m_{ll}-m_Z|$ [GeV]  &   & $>20$ (only SF)\\
$n_{b-jets}$                         & \multicolumn{2}{c}{$\geq1$}\\
$\Delta\phi_{\rm boost}$ [rad]                           &\multicolumn{2}{c}{$<1.5$}\\
 $E_{\mathrm{T}}^{\mathrm{miss}}$ significance                           &\multicolumn{2}{c}{$>12$}\\
$m_{T2}$ [GeV]                   & \multicolumn{2}{c}{$>110$}\\
\end{tabular}
\end{center}
\end{table}

\begin{figure*}[tb]
\begin{center}
      \includegraphics[width=1.2\columnwidth]{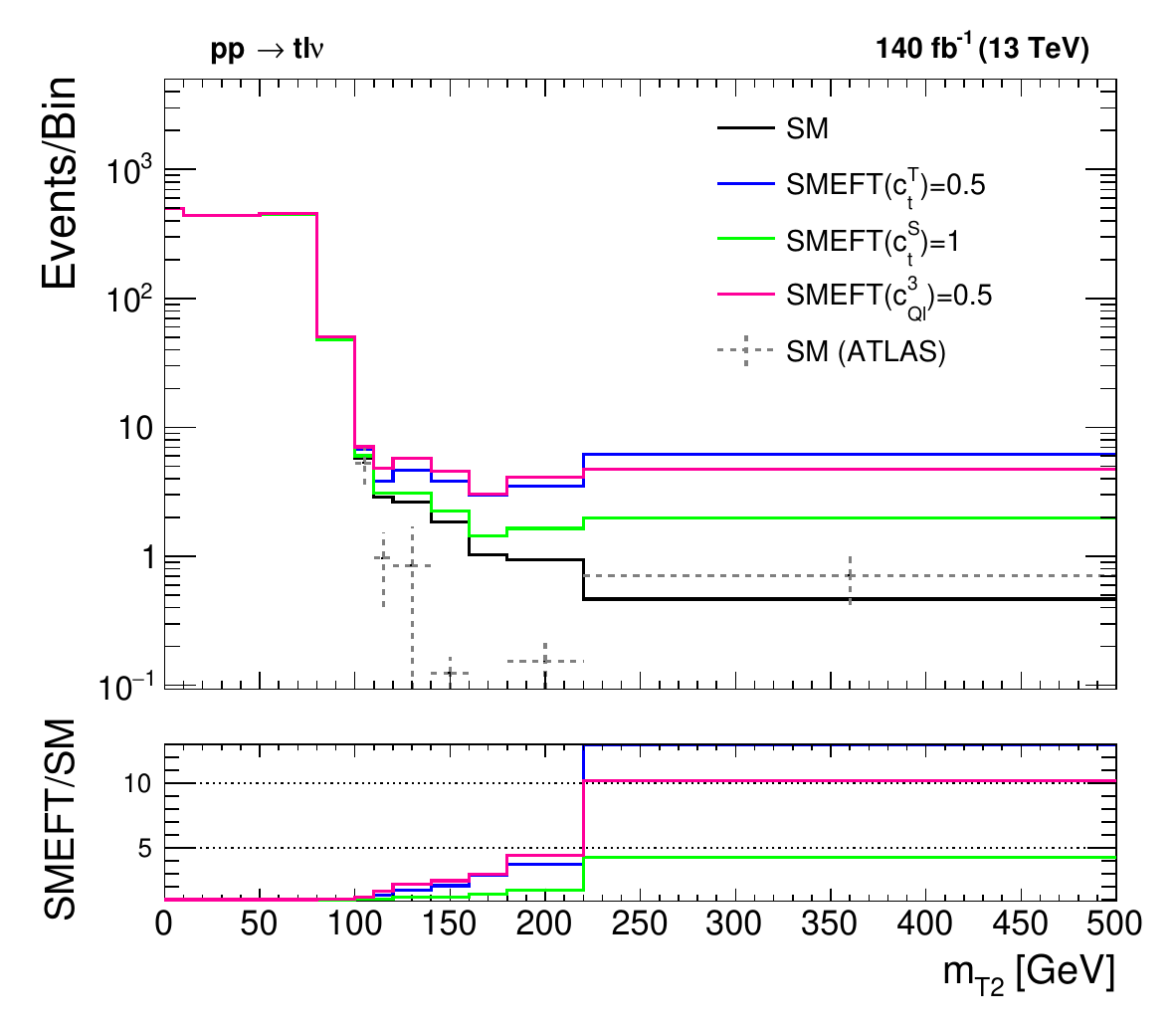}
    \caption{Distribution of the  $m_{T2}$ observable for the pp$\rightarrow$t$l\nu$ process with the two-body selection. Events with same flavor and different flavor leptons are combined. The ATLAS prediction together with total uncertainties is shown in the gray dotted histogram. The solid line histograms show the SM and SMEFT predictions without cutting on the $m_{T2}$  variable.      
    \label{fig:ATLAScomparison}}
    \end{center}
\end{figure*}

To reinterpret the results of the ATLAS search, simulated signal events are passed to \textsc{Delphes 3} for modeling the response of the ATLAS detector \cite{deFavereau:2013fsa}.
The ``two-body selection'' requirements are applied to the  \textsc{Delphes} reconstructed objects. 
It is not trivial to extract precise  object-based $E_{\mathrm{T}}^{\mathrm{miss}}$ significance from the   \textsc{Delphes} reconstructed objects.
Therefore, we approximate the object-based $E_{\mathrm{T}}^{\mathrm{miss}}$ significance by an event-based $E_{\mathrm{T}}^{\mathrm{miss}}$ significance ($\cal{S}$) as:
\begin{equation*}
\label{eq:MET}
\cal{S}=\frac{\mathrm{E}_{\mathrm{T}}^{\mathrm{miss}}}{\sqrt{\mathrm{H}}_{\mathrm{T}}}\,
\end{equation*}
where $H_T$  is the sum of the transverse momenta of visible particles  \cite{ATLAS:2018uid}.  
The cut value for the event-based $E_{\mathrm{T}}^{\mathrm{miss}}$ significance was chosen to obtain a similar SM prediction for the pp$\rightarrow$tW process in the t$\Bar{\rm t}$ validation region reported by the ATLAS collaboration (see Table 7 in Ref \cite{ATLAS:2021hza}).  The  t$\Bar{\rm t}$ validation region has exactly the same selection requirements as the ``two-body selection'' except for the $m_{T2}$ variable which is restricted to [100, 110] GeV. 
We found that cutting on the event-based $E_{\mathrm{T}}^{\mathrm{miss}}$ significance at 9.6 GeV$^{1/2}$ for our Delphes simulated sample could reproduce the reported ATLAS prediction for the pp$\rightarrow$tW process in the  t$\Bar{\rm t}$ validation region. Therefore we replace the cut  of $E_{\mathrm{T}}^{\mathrm{miss}}$ significance $>12$ GeV  with the   $\cal{S}>$9.6 GeV$^{1/2}$ requirement in the ``two-body selection''.

\begin{table*}[htb]
\caption{The observed $2\sigma$ confidence intervals for the \cQla ,  \ctlS, and \ctlT obtained in Refs.   \cite{CMS:2023xyc} and \cite{Stolarski:2020cvf} compared to 95\% confidence level observed limits obtained in this analysis. }
\begin{center}
\begin{tabular}{lcccl}
\hline
Process                      & \multicolumn{3}{c}{$2\sigma$ CI or 95\% CL limits [TeV$^{-2}$] }                                                           & Ref.                       \\ 
pp$\rightarrow$                      &           $ \cQla $  & $ \ctlS$  &                   $ \ctlT$                                 &                        \\ \hline
 t$\Bar{\rm t}\ell\ell$ and t$\Bar{\rm t}\ell\nu$              &  [-2.84, 2.55]  & [-2.60, 2.62] &   [ 0.37, 0.37] &   \cite{CMS:2023xyc} \\ 
 t$l\nu$             &[-10,10]       & [-20,20] & [-5,5] &  \cite{Stolarski:2020cvf} \\ 
 t$l\nu$          &[-0.93,0.46]       & [-1.76,1.76] & [-0.45,0.45]   &  this paper \\ 
\end{tabular}
\label{tab:resultsCom}
\end{center}
\end{table*}

In Figure \ref{fig:ATLAScomparison},  the $m_{T2}$ distribution predicted by  the ATLAS collaboration for the SM  pp$\rightarrow$tW process in the signal region is compared to the SMEFT predictions via the Delphes simulated sample. There is an acceptable agreement between our simulations at the SM point of WC phase space and the ATLAS predictions.  Although the ATLAS prediction for the SM tW process  in the tail of the $m_{T2}$ distribution has large statistical uncertainties, the agreement between our prediction and the ATLAS prediction for the SM tW yield in the last bin of the $m_{T2}$ distribution, which is the most sensitive bin, is very good. 
 In addition to the SM prediction, SMEFT predictions for the pp$\rightarrow$t$l\nu$ process in the presence of the two-quark-two-lepton operators are shown. 
 The ratio plots in Figures \ref{fig:Gen} and \ref{fig:ATLAScomparison} show similar deviations with respect to the SM prediction at the generator and reconstructed levels.  This confirms that the ``two-body selection'' is proper for probing the two-quark-two-lepton operators at the reconstruction level.

\section{Results}
No excess in data over the SM prediction is observed  by the ATLAS Collaboration~\cite{ATLAS:2021hza}. The results are presented in terms of model independent 95\% confidence levels (CL) on the number of signal events using the CLs method \cite{Read:2002hq}.   
The upper limits are derived for the  different flavor and same flavor signal regions combined. 
Multiple upper limits are calculated using event yields from the following cuts on the lower edge of the range for the  $m_{T2}$ variable, with the upper end of the range at infinity: $m_{T2}>$ 110, 120, 140, 160,  180, 200, or 220 GeV.  
We use these upper limits to constraint the two-quark-two-lepton WCs affecting t$\ell\nu$ production.

\begin{figure*}[htb]
\begin{center}
      \includegraphics[width=1.2\columnwidth]{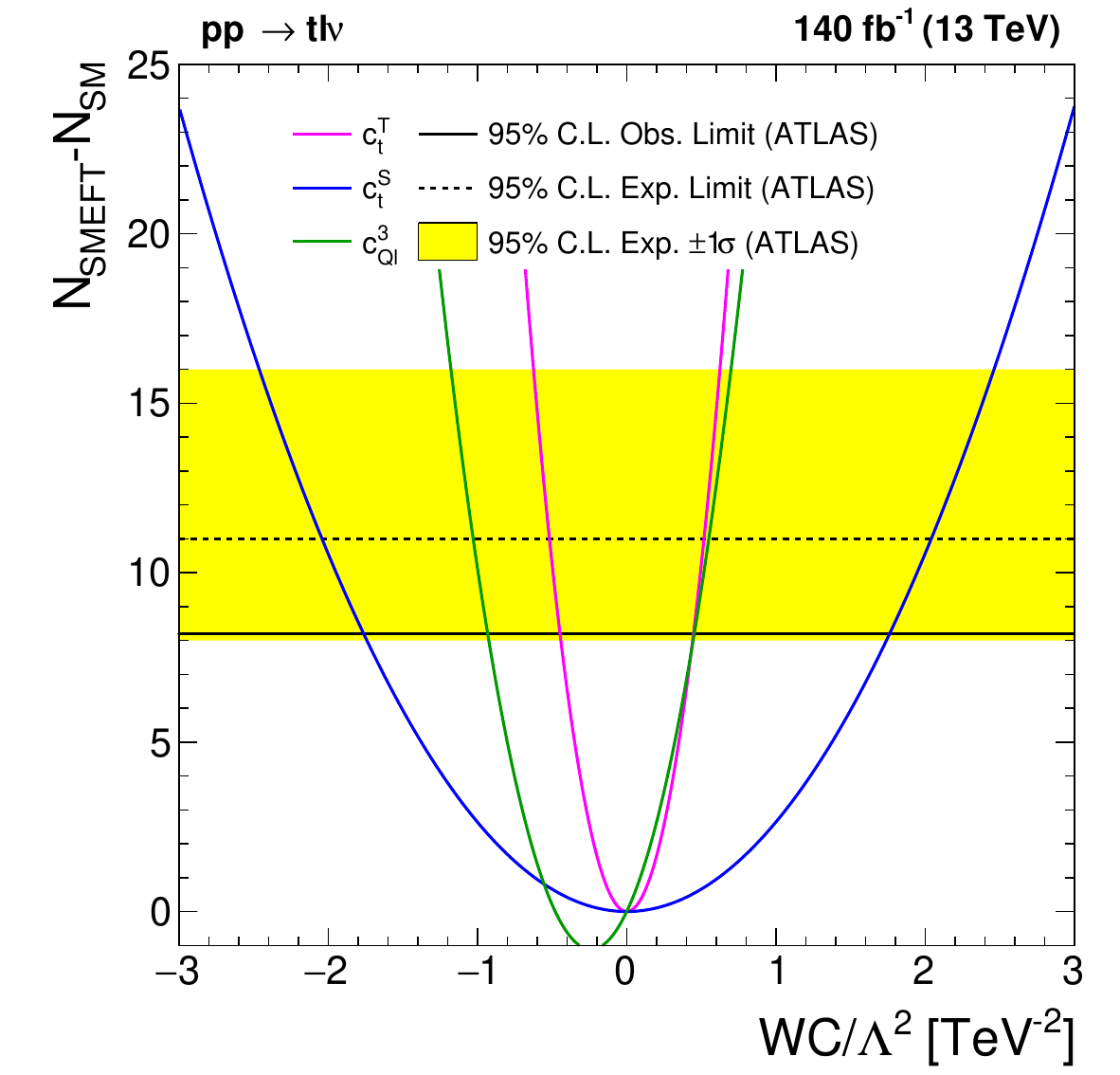}
    \caption{Difference between the SMEFT and SM predictions for the number of t$l\nu$ events  in 140 fb$^{-1}$ of data as a function of the two-quark-two-lepton WCs.  Events are selected using two-body selection requirements with $m_{T2}$=[160, $\infty$] GeV. 
    The observed and expected upper limits on the number of signal events obtained in  two-body selection region with $m_{T2}$=[160, $\infty$] GeV by the ATLAS Collaboration are  shown with solid and dashed lines, respectively. 
    \label{fig:results}}
    \end{center}
\end{figure*}

It is more sensitive to include both the yield and shape variations in multiple bins of the $m_{T2}$ distribution because of the SMEFT effects.
However, because the SMEFT signal shapes are different from the top squark signal shape we are constrained to consider on the event yield variation in one cumulative bin in order to use the model dependent limits in Ref ~\cite{ATLAS:2021hza}.
Based on the reported expected limits, we found that the tightest constraints are obtained in the range where $m_{T2}$ $\in$ [160, $\infty$] GeV. In Figure \ref{fig:results} the difference between the SMEFT  and  SM yield predictions for the pp$\rightarrow$t$\ell\nu$ process in the range $m_{T2}\in$ [160, $\infty$] GeV are shown as a function of individual two-quark-two-lepton WCs. The 95\% CL observed and expected  upper limits on the number of signal events in the same range of $m_{T2}$, reported by the ATLAS Collaboration,  are also shown. 
The constraints on the two-quark-two-lepton  WCs are determined by finding where the quadratic curves cross the 95\% CL upper limit lines. In Table \ref{tab:resultsCom}, the 95\% CL upper limits obtained in this analysis  by using the $m_{T2}$ observable are listed. 
In addition, our results are compared to the limits obtained from analyzing  the M$_T$($ \ell\ell \nu \nu b$) observable \cite{Stolarski:2020cvf} and to those extracted from events in the three lepton final states  \cite{CMS:2023xyc}. 
The obtained limit on the $ \cQla $ operator is the most stringent reported on this WC to date.
Constraints on the $ \ctlS$  and  $ \ctlT$    couplings are comparable to those obtained in the three lepton final state. 
Therefore, combining  dilepton  and three lepton final states in a global fit can improve the results on top of the individual constraints.

\section{Summary}
In this work, we have investigated the possibility of probing the top quark related two-quark-two-lepton SMEFT operators in the   production of a single top quark in association with a lepton-neutrino pair at the LHC. We showed that the pp$\rightarrow$t$l\nu$ process is very sensitive to the \qq{3}{lq}{}, \qq{1}{lequ}{}, and \qq{3}{lequ}{} operators in the tail of the $m_{T2}$ distribution.  To make a realistic evaluation of the  power of the proposed phase space for constraining these SMEFT operators, we reinterpreted the results of an ATLAS search in a very similar phase space.
We set strong 95\%  confidence level limits on the Wilson coefficients of the mentioned operators.  
The obtained bound on the $ \cQla $ operator is around four times tighter than the latest reported bounds to date.
Consequently, we advocate for including the $m_{T2}$ variable as analyzed in this paper in future SMEFT searches and global SMFFT fits as a probe of this specific corner of phase space.


\section*{Acknowledgments}
R.G. would like to thank Hamzeh Khanpour for useful input. 
	
	\addcontentsline{toc}{section}{References}
	\bibliographystyle{JHEP}
	\bibliography{paper}

\end{document}